\documentstyle[12pt,psfig]{article}
\let\section=\subsection     \let\subsection=\subsubsection                %%
\begin{document}
\begin{center}
   {\large \bf STRANGE PHASES IN NEUTRON STARS}\\[5mm]
   J.~SCHAFFNER-BIELICH and N.~K.~GLENDENNING \\[5mm]
   {\small \it  Nuclear Science Division, 
    Lawrence Berkeley National Laboratory \\
    University of California, Berkeley 94720, USA \\[8mm] }
\end{center}

\begin{abstract}\noindent
The equation of state and the properties of neutron stars are studied for
a phase transition to a charged kaon condensate. 
We study the mixed phase by using Gibbs condition with comparison to the
hitherto applied Maxwell construction.
Implications for kaon condensation and for the mass-radius relation of
condensed neutron stars are examined.
\end{abstract}

The high density behavior of the equation of state of nuclear matter is 
still quite unknown and has led to speculations about the appearance of new
phases. At sufficiently high density, there should be a phase transition to a
quark plasma. In the following, we will discuss the onset of kaon condensation
in neutron matter 
which might happen at lower density. Kaon condensation in connection
with $\beta$-stable matter has already been studied in \cite{Glen85}. It
received considerable attention after the work of \cite{KN86} which
included strong in-medium effects for the kaons. Once the effective kaon energy
hits the electrochemical potential, neutrons can convert to protons and $K^-$.
Using chiral perturbation theory, this happens at $(3-4)\rho_0$ \cite{Brown92}.
As the kaon is a boson and condenses, the equation of state is considerably
softened and the maximum mass of a neutron star is lowered to $1.5 M_\odot$
\cite{Thorsson94}. This is turn provides a scenario for low mass black holes as
proposed by Bethe and Brown \cite{BB94}. As a side remark, it was found
before the kaon condensation scenario was introduced 
that the less spectacular appearance of hyperons results also in a lower
maximum mass of neutron stars \cite{Glen85}.

The essential ingredient for kaon condensation is the lowering of the
effective mass of the kaon in the medium. One knows from kaon-nucleon
scattering that the s-wave $K^+$N scattering is repulsive. The low density
theorem then states that the optical potential in the nuclear medium is then
also repulsive and about +30 MeV at normal nuclear density.
Surprisingly, also the scattering for the antiparticle, the $K^-$, shows
repulsion which is due to the appearance of the $\Lambda(1405)$ resonance just
below threshold \cite{Siegel88}. 
A recent analysis of $K^-$ atoms suggest that the optical potential of the
$K^-$ can be as deeply attractive as --200 MeV at normal nuclear density
\cite{Fried93}. A coupled channel calculation by Koch \cite{Koch94}
including in-medium effects actually resolved this: as the repulsive 
$\Lambda(1405)$ mode
is shifted up in the medium due to Pauli-blocking effects, it leaves only an
attractive mode with an optical potential of about --100 MeV at $\rho_0$.

Hence, the $K^-$ feels attraction in dense nuclear matter and might condense.
Using a Maxwell construction for the phase transition, it was found that the
mass-radius relation changes considerably for a kaon condensed star
\cite{Thorsson94}. Nevertheless, there are two chemical potentials for neutron
star matter, the baryochemical potential and the electrochemical potential, as
baryon number and charge are conserved quantities. If the phases are in
chemical and mechanical equilibrium then the pressure and all chemical
potentials involved should be the same in the two phases:
\begin{equation}
p^{\rm I} = p^{\rm II} \qquad \mu_B^{\rm I} = \mu_B^{\rm II}  \qquad  
\mu_e^{\rm I} = \mu_e^{\rm II} 
\quad .
\end{equation}
This is just Gibbs general condition for two phases in equilibrium.
A Maxwell construction can assure that only one chemical potential is common to
the two phases and  
is therefore not applicable to neutron star matter \cite{Glen92}. 
One has an additional degree of freedom to maximize the pressure of the system:
charge. According to Gibbs, there exist now three possible solutions: 
the pure nucleon phase with zero charge, the pure kaon condensed phase with
zero charge, and the mixed phase where the two phases 
have finite charge densities but neutralize each other.
Geometric structures will appear with different local charges. 

We model the onset to kaon condensation by using the relativistic mean-field
model and the following Lagrangian
for the kaon-nucleon interaction
\begin{equation}
{\cal L}_K = D^*_\mu \bar K D^\mu K - {m^*_K}^2 \bar KK
\end{equation}
with a minimal coupling to scalar and vector fields
\begin{eqnarray}
D_\mu &=& \partial_\mu + i g_{\omega K} V_\mu + i g_{\rho K} \tau R_\mu \cr
m^*_K &=& m_K + g_{\sigma K} \sigma 
\quad .
\end{eqnarray}
The vector coupling constants are fixed by simple quark counting rules,
$g_{\omega K} = g_{\omega N}/3$ and $g_{\rho K} = g_{\rho N}$.
The scalar coupling constant is fixed to the optical potential of the $K^-$ at
$\rho_0$. 

\begin{figure}
\begin{center}
\leavevmode
\psfig{file=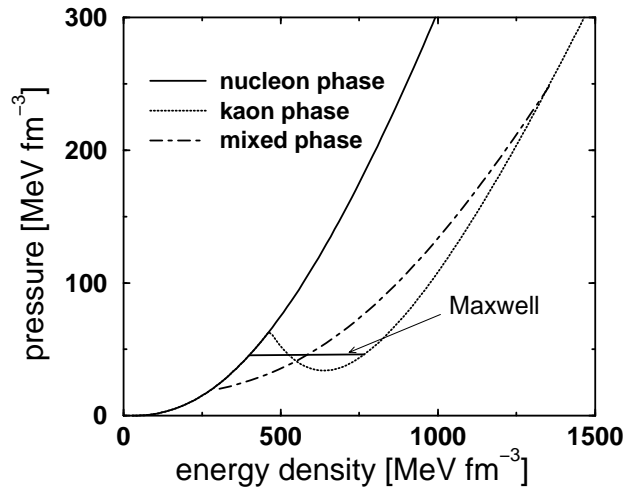,height=3in}
\end{center}
\vspace*{-1cm}
\caption{The equation of state for neutron matter with a kaon condensed phase
using a Maxwell construction compared to using Gibbs conditions.}
\label{fig:eosmg}
\end{figure}

Figure~\ref{fig:eosmg} 
shows the equation of state of neutron matter with a kaon
condensed phase. All three solutions are plotted for an optical potential of
the $K^-$ of $U_K=-140$ MeV. 
For comparison, the horizontal
line shows the case when using an unphysical
Maxwell construction that assumes equality of $\mu_B$ in both phases but not of
$\mu_e$ resulting in a constant pressure.  
The kaon condensed phase shows some instability, i.e.\ a negative curvature,
close to the onset of the phase transition point.
The mixed phase spans over a
wide range of density and starts at much lower density  
than the pure kaon condensed phase. This solution gives then a continuously
growing equation of state without any instability.
The equation of state is considerably smoothened compared to the case of a
Maxwell construction. 
It appears much like the equation of state of a second order phase transition
with the important difference, however, that there is a density range
for which a mixture of two distinct phases occurs.

The charge is distributed between the two phases in the mixed phase. 
At the beginning of the mixed phase, the bubbles of the new phase are highly
negatively charge while occupying only a small amount of the total volume.
At higher density, the two phases have just the opposite charge and share
equally the available volume. At the end of the phase transition, the situation
is reversed and the nucleon phase is highly positively charged while being
only in a small fraction of the total volume.
Note, that the asymmetry energy of nuclear matter is the driving force 
so that there is a 
positive charge in the nucleon phase while the negatively charged kaon in the
second phase allows also for a more isospin symmetric system.

\begin{figure}
\begin{center}
\leavevmode
\psfig{file=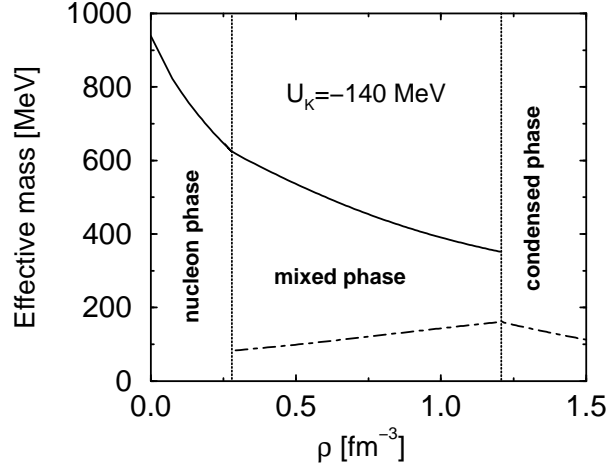,height=3in}
\end{center}
\vspace*{-1cm}
\caption{The effective mass of the nucleon in the various phases of neutron
star matter.} 
\label{fig:efmass}
\end{figure}

There are nucleons in both phases. But they do not mix with each other, because
they have different in-medium properties. The presence of the kaon shifts the
effective mass of the nucleon to lower values as there are different field
configurations in this phase. This is visualized in 
Figure~\ref{fig:efmass}. The effective mass of the nucleon is lowered at higher
density. In the mixed phase starting at $3\rho_0$, 
the effective masses of the nucleons on the two
phases are different from each other. In the case shown, the effective mass in
the kaon condensed phase is about 300 MeV lower than in the nucleon phase.
At the end of the mixed phase at $6.5\rho_0$, 
the solution for the kaon condensed phase
remains and the effective mass of the nucleon stays around 200 MeV.

\begin{figure}
\begin{center}
\leavevmode
\psfig{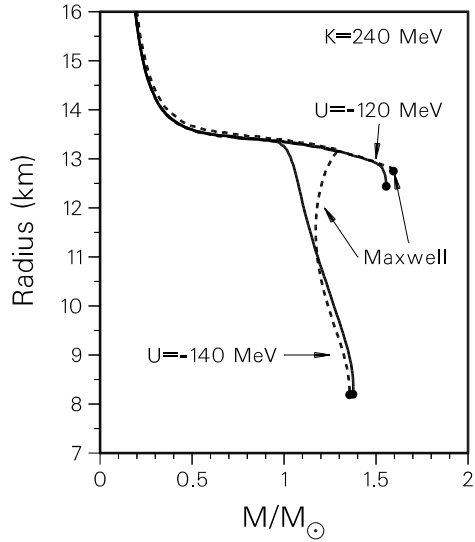}
\end{center}
\caption{The mass-radius relation of a neutron star with a kaon condensed
phase.} 
\label{fig:mrkaon}
\end{figure}

Finally, we discuss the mass-radius relation for a kaon condensed neutron star
in Figure~\ref{fig:mrkaon} for two choices of the optical potential of
the $K^-$ at $\rho_0$. The maximum mass is between $(1.6-1.4)M_\odot$ and does
not change very much for 
an optical potential between $U_K=-120$ and  $-140$ MeV. 
The main difference is the
minimum radius reached in the two cases: for the stronger potential the minimum
radius is $R=8$ km while it is still above $R=12$ km for the weaker potential.
The dotted lines are the case when using a Maxwell construction.
For some intermediate range, this curve shows a mechanical instability due to
the constant pressure. For very high density, the dotted line is then close to
the Gibbs construction as the pure kaon condensed phase is reached in the
interior of the neutron star where the
two constructions coincide. According to \cite{Koch94}, the weaker potential
is the more plausible.

In summary, kaon condensation results in an equation of state without a
constant pressure zone when applying Gibbs criteria. A mixed phase can exist
with a rigid structure with different local charges. The mass-radius relation
is considerably smoothened compared to a Maxwell construction.

The present calculation ignores two major points. Finite size effects of the
structures in the mixed phase as surface tension and Coulomb energy will 
change the appearance and disappearance of the mixed phase. 
The surface tension can be calculated within the model in semi-infinite matter
in a selfconsistent manner which we will pursue in the future. 
Secondly, the appearance of hyperons will shift the onset of kaon condensation
to a higher density \cite{Knorren95a,Knorren95b} 
or can even lead to their disappearance
\cite{Glen85,SM96}. 
The inclusion of hyperons will be studied in forthcoming work.

\subsection*{Acknowledgments}

J. S.-B. acknowledges support
by the Alexander-von-Humboldt Stiftung with a Feodor-Lynen fellowship.
This work is supported in part by 
the Director, Office of Energy Research,
Office of High Energy and Nuclear Physics, Nuclear Physics Division of the
U.S. Department of Energy under Contract No.\ DE-AC03-76SF00098.

\end{document}